\documentclass[12pt]{article}
\usepackage{graphicx,epsfig,color}
\renewcommand{\theequation}{\arabic{section}.\arabic{equation}}
\newcommand{\bc}{\begin{center}}
\newcommand{\ec}{\end{center}}
\newcommand{\be}{\begin{equation}}
\newcommand{\ee}{\end{equation}}
\newcommand{\bea}{\begin{eqnarray}}
\newcommand{\eea}{\end{eqnarray}}
\newcommand{\ba}{\begin{array}}
\newcommand{\ea}{\end{array}}
\newcommand{\lb}{\label}
\newcommand{\rf}{\ref}
\newcommand{\bfg}{\begin{figure}[htbp]}
\newcommand{\efg}{\end{figure}}
\newcommand{\pr}{Phys. Rev. }
\newcommand{\np}{Nucl. Phys. }
\newcommand{\prl}{Phys. Rev. Lett. }
\newcommand{\prp}{Phys. Rep. }

\newcommand{\pl}{Phys. Lett. }

\newcommand{\jp}{J. Phys. }
\newcommand{\rmp}{Rev. Mod. Phys. }

\begin{document}

\vspace*{1. cm}
\bc
{\large \textbf{Spectral properties of the gauge invariant\\
\vspace{0.25 cm}
quark Green's function in two-dimensional QCD}}\\
\vspace{1 cm}
H. Sazdjian\\
\textit{Institut de Physique Nucl\'eaire, CNRS/IN2P3,\\
Universit\'e Paris-Sud 11, F-91405 Orsay, France\\
E-mail: sazdjian@ipno.in2p3.fr}
\ec
\par
\renewcommand{\thefootnote}{\fnsymbol{footnote}}
\vspace{0.75 cm}

\bc
{\large Abstract}
\ec
\par
The gauge invariant quark Green's function with a path-ordered phase
factor along a straight-line is studied in two-dimensional QCD in
the large-$N_c$ limit by means of an exact integro\-differential
equation. Its spectral functions are analytically determined. They are 
infra-red finite and lie on the positive real axis of the complex plane 
of the momentum squared variable, corresponding to momenta in the forward 
light cone. Their singularities are represented by an infinite number of
threshold type branch points with power-law $-3/2$, starting at 
positive mass values, characterized by an integer number $n$ and 
increasing with $n$. The analytic expression of the Green's function
for all momenta is presented. The appearance of strong threshold 
singularities is suggestive of the fact that quarks could not be observed 
as asymptotic states.  
\par
\vspace{0.5 cm}
\par

\newpage

\section{Introduction} \lb{s1}
\setcounter{equation}{0}

Conventional quark Green's functions suffer from lack of
gauge invariance. In particular, no physical conclusion can be 
drawn from any of their individual properties. Their role is
rather limited to lead, together with other Green's functions, 
to the evaluation of gauge invariant observable quantities, like
bound state masses, current matrix elements, hadronic form factors,
scattering amplitudes, etc. This is not the case of the gauge invariant 
quark Green's functions (GIQGF), constructed with the aid of the gluon 
field path-ordered phase factors \cite{m,nm}. Their properties should 
provide a more stable basis for the investigation of the physical  
characteristics of the underlying quark and gluon fields. Nevertheless,
because of the mathematical complexity of the corresponding composite
objects, not much has been known up to now about them. 
\par
Recently, the present author could obtain, using a representation for 
the quark propagator in an external gluon field that generalizes the 
one introduced in the nonrelativistic case \cite{ef}, exact 
integrodifferential equations for the the two-point GIQGFs (2PGIQGF), 
in which the paths of the phase factors are of the skew-polygonal type 
\cite{s}. It turns out that such GIQGFs form a closed set with respect 
to the above equations. The kernels of these equations are represented 
by functional derivatives of Wilson loop averages \cite{w,p,mm,mgd,mk}. 
Designating by $S_{(1)}(x,x')\equiv S(x,x')$ the 2PGIQGF with a path 
made of one straight line joining the quark to the antiquark and by 
$S_{(n)}(x,x';y_{n-1},\ldots,y_1)$ ($n\geq 2$) the 2PGIQGF
with a path made of $n$ segments and $n-1$ junction points 
$y_1,y_2,\ldots,y_{n-1}$ between the segments, the equation satisfied
by $S$ has the following structure:
\bea \lb{1e1}
& &(i\gamma.\partial_{(x)}-m)S(x,x')=i\delta^4(x-x')
+i\gamma^{\mu}\,\Big\{K_{2\mu}(x',x,y_1)\,S_{(2)}(y_1,x';x)\nonumber \\
& &\ \ \ \ +\sum_{i=3}^{\infty}K_{i\mu}(x',x,y_1,\ldots,y_{i-1})\,
S_{(i)}(y_{i-1},x';x,y_1,\ldots,y_{i-2})\Big\}.
\eea
[$m$ is the quark mass parameter. Integrations on the intermediate 
variables $y$ are implicit.]
The kernels $K_{n}$ ($n=2,\ldots,\infty$) contain logarithms of Wilson 
loop averages along the contour of the $(n+1)$-sided skew-polygon with
at most $n$ functional derivatives along the contour and $(n-1)$ GIQGFs
$S$ along the segments of the contour. The 2PGIQGFs $S_{(n)}$ ($n\geq 2$)
are themselves related to the lowest-order 2PGIQGF $S$ through functional
relations involving mainly the Wilson loop average along the 
$(n+1)$-sided skew-polygon. It is then conceivable that, by eliminating
the $S_{(n)}$s in favor of $S$, one might reach a closed form equation 
for $S$. 
\par
What kind of information could provide us the analysis of the 
2PGIQGF $S$? The status of that Green's function is rather intricate.
If colored objects are confined and if singularities of Green's
functions result from insertions of complete sets of physical states,
which are here composed of hadronic states (color singlets), then it is 
not possible to cut the line joining the quark to the antiquark and 
saturate it with physical states, the contribution of the latter being
identically zero. This situation would suggest that the above
Green's function is free of singularities. Nevertheless, the equation
that it satisfies and which is derived from the QCD lagrangian signals
the presence of singularities, generated by the free quark propagator.
This paradoxical situation is solved only with the acceptance that
the building blocks of the theory, which are the quark and gluon fields,
continue forming a complete set of states with positive energies and
could be used for any saturation scheme of intermediate states. It is
then up to the theory to indicate us through the solutions of the
equations of motion how these singularities combine to form the
complete contributions. For sectors of colored states, we should
expect to find singularities, whenever they are present, which would
be stronger than the observable ones. Therefore, the determination
of the 2PGIQGF would provide us a direct test of the confinement 
mechanism in the colored sector of quarks.
\par
Saturating the intermediate states of the 2PGIQGF $S$ with
quarks and gluons with positive energies, one arrives at a generalized
form of the K\"allen-Lehmann representation \cite{s}:
\be \lb{1e2}
S(p)=i\int_0^{\infty}ds\,\sum_{n=1}^{\infty}\,
\frac{\big[\,\gamma.p\,\rho_1^{(n)}(s)+\rho_0^{(n)}(s)\,\big]}
{(p^2-s+i\varepsilon)^n}.
\ee
[$S(p)$ is the Fourier transform of $S(x,x')=S(x-x')$.] The increasing 
powers of the denominators come from the increasing number of gluon fields
in the series expansion of the phase factor in terms of the coupling 
constant. Formally, one might reduce, by integrations by parts, all the
denominators to a single one, but the possible presence of strong threshold
singularities might stop the operation at some power. In the latter case, 
another alternative would be the extraction out of the integral of 
derivation operators. 
\par
There might also exist other extreme cases. For instance, the 
singularities of $S$, as provided by the solution of Eq. (\rf{1e1}),
might lie in unphysical regions, like the complex plane or the negative
axis of $p^2$. Or, the summation of the series (\rf{1e2}) might lead to
new singularities in unphysical regions. We shall, however, continue
exploring the optimistic scenario, according to which the series of
singularities (\rf{1e2}) are the only type of singularities of $S$ such
that their sum remains on the physical timelike axis of $p^2$.
\par
This paper is devoted to the resolution of Eq. (\rf{1e1}) in 
two-dimensional QCD in the large-$N_c$ limit \cite{th2,th3,ccg} . That
theory displays the main features of confinement of the four-dimensional 
case with the additional simplification that asymptotic freedom
is realized here in a rather trivial way, since the theory is 
superrenormalizable. Furthermore, many aspects of two-dimensional theories 
are often exactly solvable and in this case one might hope to  have an 
unambiguous check of the physical implications of Eq. (\rf{1e1}) for the 
spectral properties of the 2PGIQGF.
\par
The main results of the present investigation can be summarized as
follows. The series of spectral functions of the vector and scalar   
types of Eq. (\rf{1e2}) can actually be grouped into two single
spectral functions. The singularities of the latter can be
analytically determined. They are represented, by an infinite number of 
threshold type branch points with power-law $-3/2$, starting at 
positive mass values characterized by an integer number $n$ and increasing 
with $n$. This situation is qualitatively unchanged in the limit of 
massless quarks. The strong thereshold singularities, together with the 
absence of pole terms, are suggestive of the fact that quarks could not 
be observed as physical asymptotic states.     
\par
The plan of the paper is the following. In Sec. \rf{s2}, we consider 
two-dimensional QCD in the large-$N_c$ limit. In Sec. \rf{s3}, the Green's 
function is considered in $x$-space and its properties are determined.
In Sec. \rf{s4}, the instantaneous limit of the Green's function in
momentum space is taken. In Sec. \rf{s5}, the analyticity properties of
the Green's function in the instantaneous limit are studied and the
spectral functions are reconstructed. Sec. \rf{s6} is devoted to a brief 
application to the calculation of the quark condensate. Conclusion
follows in Sec. \rf{s7}. An appendix is devoted to the study of the 
properties of the threshold masses and to their determination. 
\par

\section{Two-dimensional QCD in the large-$N_c$ limit} \lb{s2}
\setcounter{equation}{0}

On technical grounds, considerable simplification occurs in Eq. (\rf{1e1})
in two-dimensional QCD at large $N_c$. Wilson loop averages are now 
exponential functionals of the areas of the surfaces lying inside the 
contours \cite{kkk,br}. First, for simple contours, made of junctions of 
simpler ones, the Wilson loop average factorizes into the product of the 
averages over the simpler contours. This feature leads to the disappearance 
in the kernels of Eq. (\rf{1e1}) of nested-type diagrams. Second, crossed 
type diagrams as well as quark loop contributions disappear. Third, the 
second-order functional derivative of the logarithm of the Wilson loop 
average is a (two-dimensional) delta-function. As a consequence, since
each derivation acts on a different segment, all functional derivatives 
of higher-order than the second vanish. The series of kernels in the 
right-hand side of Eq. (\rf{1e1}) then reduces to its first term. Equation 
(\rf{1e1}) becomes:
\be \lb{2e1}
(i\gamma.\partial_{(x)}-m)S(x-x')=i\delta^2(x-x')
-i\gamma^{\mu}\left(\frac{\bar\delta^2 F_3(x',x,y)}{\bar\delta x^{\mu -}
\bar\delta y^{\alpha +}}\right)\,S(x-y)\,\gamma^{\alpha}\,
S_{(2)}(y,x';x),
\ee
where $F_3$ is the logarithm of the Wilson loop average along the 
triangle $xyx'$. The functional derivatives acting on $F_3$
represent integrated derivatives on the segments $xx'$ and $yx'$, $x'$ 
remaining fixed \cite{s}; one has:
\newpage
\bea \lb{2e2}
\frac{\bar\delta^2 F_3(x',x,y)}{\bar\delta x^{\mu -}
\bar\delta y^{\alpha +}}&=&2i\sigma(g_{\mu\alpha}g_{\nu\beta}
-g_{\mu\beta}g_{\nu\alpha})(x'-x)^{\nu}(y-x')^{\beta}\nonumber \\
& &\ \ \ \times \int_0^1 d\lambda\,d\lambda'\,\lambda\,\lambda'
\delta^2(\lambda'(y-x')+\lambda(x'-x)),
\eea
where $\sigma$ is the string tension. The $\delta$-function forces the 
point $y$ to lie on the line $x'x$. In the present theory, the 2PGIQGF 
$S_{(2)}$ is related in a simple way to $S$:
$S_{(2)}(y,x';x)=e^{F_3(x',x,y)}S(y-x')$. Integrating with respect
to $y$ and one of the $\lambda$s with an integration by parts, one
obtains the following (exact) equation:
\bea \lb{2e3}
& &(i\gamma.\partial-m)S(x)=i\delta^2(x)
-\sigma\gamma^{\mu}(g_{\mu\alpha}g_{\nu\beta}-g_{\mu\beta}g_{\nu\alpha}) 
x^{\nu}x^{\beta}\nonumber \\
& &\ \ \ \ \ \times\left[\,\int_0^1d\lambda\,\lambda^2\,S((1-\lambda)x)
\gamma^{\alpha}S(\lambda x)
+\int_1^{\infty}d\xi\,S((1-\xi)x)\gamma^{\alpha}S(\xi x)\,\right]. 
\eea
The first integral represents contributions when $y$ is lying inside the
segment $x'x$, while the second one contributions from outside the
segment $x'x$ in the direction $x'x$. 
\par

\section{The quark Green's function} \lb{s3}
\setcounter{equation}{0}

Let us first observe that attempting to develop Eq. (\rf{2e3})
into an iterative series with respect to the free propagator
would lead to a series of terms with increasing singularities at
the quark mass $m$. Replacing in the interaction dependent part of 
Eq. (\rf{2e3}) $S$ by the free propagator $S_0$, produces a term 
that behaves at the quark mass threshold as $(p^2-m^2)^{-5/2}$, 
which, in turn leads to stronger singularities at higher orders. 
Therefore, the existence of a nonperturbative solution, with 
possibly stable properties concerning its singularities, would 
represent the only issue to have access to the structure of the 
complete solution. It turns out that Eq. (\rf{2e3}) does have such 
a solution satisfying the required boundary conditions and expressible 
in analytic form. A complete understanding of the spectral properties 
of the Green's function $S$ requires an analysis in momentume space. 
We defer that study to Secs. \rf{s4} and \rf{s5}. In this section, we 
shall display the solution and shall check in $x$-space that it 
satisfies Eq. (\rf{2e3}). 
\par  
Decomposing the momentum space 2PGIQGF $S(p)$ along vector and scalar 
invariants $F_1$ and $F_0$,
\be \lb{3e1}
S(p)=\gamma.pF_1(p^2)+F_0(p^2),
\ee
the solutions for $F_1$ and $F_0$, for complex $p^2$, are: 
\bea
\lb{3e2}
& &F_1(p^2)=-i\frac{\pi}{2\sigma}\,\sum_{n=1}^{\infty}\,
b_n\,\frac{1}{(M_n^2-p^2)^{3/2}},\\
\lb{3e3}
& &F_0(p^2)=i\frac{\pi}{2\sigma}\,\sum_{n=1}^{\infty}\,
(-1)^nb_n\,\frac{M_n}{(M_n^2-p^2)^{3/2}},
\eea
where the coefficients $b_n$ are parametrized as
\be \lb{3e4a}
b_n=\frac{\sigma^2}{M_n+\overline m_n+(-1)^nm}.
\ee
The mass parameters $M_n$ have real positive values, greater than the
free quark mass $m$, increasing with $n$, while the parameters 
$\overline m_n$ generally have real positive values decreasing, for a 
fixed parity of the index $n$, with $n$. They are defined through an 
infinite set of algebraic equations which will be displayed below. 
Their asymptotic values, for large $n$, such that $n\gg m^2/(\sigma\pi)$, 
are:
\be \lb{3e4}
M_n^2\simeq \sigma\pi n,\ \ \ \ \ \ \ \ \  
\overline m_n\simeq \frac{\sigma}{\pi}\frac{1}{M_n}.
\ee
\par
The functions $(M_n^2-p^2)^{3/2}$ are defined with cuts starting from
their branch points and going to $+\infty$ on the real axis; they are
real below their branch points down to $-\infty$ on the real axis.
\par
From Eqs. (\rf{3e2}) and (\rf{3e3}) we deduce that the singularities
of the functions $F_1$ and $F_0$ are located on the positive real axis 
of $p^2$ and represented by an infinite number of branch points or
thresholds with stronger singularities than simple poles. That feature
is an indication that the resulting singularities are not of the
observable type.
\par 
The first threshold is located at $M_1^2$, which is positive and larger 
than the free quark mass squared $m^2$. For massless quarks, $M_1$ 
remains positive. 
\par
To analyze the solution in $x$-space, we define the Fourier transform
of the decomposition (\rf{3e1}):
\be \lb{3e5}
S(x)=i\gamma.\partial F_1(r)+F_0(r),\ \ \ \ \ r=\sqrt{-x^2}.
\ee
We introduce modified functions $\widetilde F_1(r)$ and 
$\widetilde F_0(r)$ such that
\be \lb{3e6}
S(x)=\frac{1}{2\pi}\Big(\frac{i\gamma.x}{r}\widetilde F_1(r)
+\widetilde F_0(r)\Big),\ \ \ \ 
\widetilde F_1(r)=-2\pi\frac{\partial F_1(r)}{\partial r},\ \ \ \
\widetilde F_0(r)=2\pi F_0(r).
\ee
From the Fourier transforms of Eqs. (\rf{3e2}) and (\rf{3e3}), one
obtains:
\be \lb{3e7}
\widetilde F_1(r)=\frac{\pi}{2\sigma}\,\sum_{n=1}^{\infty}\,
b_n\,e^{-M_nr},\ \ \ \ \
\widetilde F_0(r)=\frac{\pi}{2\sigma}\,\sum_{n=1}^{\infty}\,
(-1)^{n+1}b_n\,e^{-M_nr}.
\ee
\par
In terms of the functions $\widetilde F_1$ and $\widetilde F_0$, 
Eq. (\rf{2e3}) decomposes into two equations (here written for 
$r\neq 0$):
\bea
\lb{3e8}
\frac{\partial}{\partial r}(r\widetilde F_1(r))
+mr\widetilde F_0(r)&=&\frac{\sigma r^3}
{2\pi}\bigg\{\,\int_0^1d\lambda\,\lambda^2\,
\Big[\widetilde F_1((1-\lambda)r)\widetilde F_1(\lambda r) 
-\widetilde F_0((1-\lambda)r)\widetilde F_0(\lambda r)\Big]\nonumber \\
& &\ \ \ -\int_1^{\infty}d\xi\,
\Big[\widetilde F_1((\xi-1)r)\widetilde F_1(\xi r)
+\widetilde F_0((\xi-1)r)\widetilde F_0(\xi r)\Big]\,\bigg\},\nonumber \\
& &\\
\lb{3e9}   
\frac{\partial}{\partial r}(\widetilde F_0(r))
+m\widetilde F_1(r)&=&\frac{\sigma r^2}
{2\pi}\bigg\{\,\int_0^1d\lambda\,\lambda^2\,
\Big[\widetilde F_1((1-\lambda)r)\widetilde F_0(\lambda r) 
-\widetilde F_0((1-\lambda)r)\widetilde F_1(\lambda r)\Big]\nonumber \\
& &\ \ \ -\int_1^{\infty}d\xi\,
\Big[\widetilde F_1((\xi-1)r)\widetilde F_0(\xi r)
+\widetilde F_0((\xi-1)r)\widetilde F_1(\xi r)\Big]\,\bigg\}.\nonumber \\
& &
\eea
Replacing in them $\widetilde F_1$ and $\widetilde F_0$ with their
expressions (\rf{3e7}) and considering each exponential function
with a given $M_n$ as an independent function for $r\neq 0$, one obtains
the following three sets of consistency conditions, valid for all 
$n=1,\ldots,\infty$, $n$ not summed:
\bea
\lb{3e10}
& &M_n=(-1)^{n+1}m+\frac{1}{2}\,\sum_{q\neq n}\,(1-(-1)^{n+q})
\frac{b_q}{(M_n-M_q)^2},\\
\lb{3e11}
& &1=-\sum_{q\neq n}\,(1-(-1)^{n+q})
\frac{b_q}{(M_n-M_q)^3},\\
\lb{3e12}
& &\sum_{q\neq n}\,(1-(-1)^{n+q})\frac{b_q}{(M_n-M_q)}
+\sum_{q}\,(1+(-1)^{n+q})\frac{b_q}{(M_n+M_q)}=0.
\eea
\par
Equations (\rf{3e10}) and (\rf{3e12}) are consequences of both of
Eqs. (\rf{3e8}) and (\rf{3e9}), while Eq. (\rf{3e11}) results from
Eq. (\rf{3e8}) only; its components play a particular role in the limit 
$r\rightarrow 0$. In that limit, the right-hand side of Eq. 
(\rf{3e11}), which is inserted in the series of the exponential 
functions, gives a vanishing contribution for antisymmetry reasons,
while its left-hand side contributes globally as a divergent constant
reproducing the delta-function of the right-hand side of Eq. (\rf{2e3}).
Equations (\rf{3e10})-(\rf{3e12}) determine completely the mass
parameters $M_n$ and $\overline m_n$, up to a global normalization 
factor fixed by the inhomogeneous part of Eq. (\rf{2e3}). 
Details of the resolution method are presented in the appendix.
Equations (\rf{3e10}) and (\rf{3e11}) mainly determine the parameters $M$.
The parameters $\overline m$ are asymptotically negligible in front of
the $M$s and they play a secondary role in the above equations. They
are mainly determined through Eqs. (\rf{3e12}). From the latter equations 
and from Eqs. (\rf{3e10}), one derives the following ones:
\be \lb{3e17}  
\overline m_n=\frac{1}{2}\sum_{q=1}^{\infty}\,
(1+(-1)^{n+q})\frac{b_q}{(M_n+M_q)^2},\ \ \ \ \ n=1,2,\ldots,
\ee
which are more appropriate for an iterative determination of the 
$\overline m$s.
\par
Finally, we study the effect of the inhomogeneous part of Eq. (\rf{2e3}).
From Eqs. (\rf{3e8}) and (\rf{3e9}) one finds that the interaction
dependent parts, represented by the right-hand sides, vanish in the 
limit $r=0$. In momentum space this means that the asymptotic behavior
of $S(p)$ is governed, for spacelike momenta, by the free propagator, as 
can be deduced from the inhomogeneous part of Eq. (\rf{2e3}). The 
asymptotic behavior of the scalar function $F_0$, however, crucially 
depends on whether the quark is massless or not \cite{pl}. 
\par
We first check the behavior of $F_1$ from Eq. (\rf{3e2}). Taking 
$|p^2|$ large and neglecting in the right-hand side the $M_n^2$s in
front of $p^2$, one finds a divergent coefficient which is an 
indication that the asymptotic behavior of $F_1$ is slower than 
$(-p^2)^{-3/2}$. To evaluate the asymptotic tail of the series, 
one can replace $b_n$ by its asymptotic behavior 
$\sigma^2/M_n\simeq \sigma^2/\sqrt{\sigma\pi n}$ and convert the series 
into an integral. One then obtains for the asymptotic behavior 
\be \lb{3e18a}
F_1(p^2)_{\stackrel{{\displaystyle =}}{p^2\rightarrow -\infty}}
\ \frac{i}{p^2},
\ee
in agreement with what is expected from the inhomogeneous part of
Eq. (\rf{2e3}). In $x$-space, $F_1(r)$ behaves for spacelike $x$ and
small $r$ as
$(1/(4\pi))\ln(1/(\sigma r^2))$ \cite{gs} and $\widetilde F_1(r)$ as 
$1/r$.
\par 
For the study of the asymptotic behavior of $F_0$, for $m\neq 0$,
one again sticks to the asymptotic tail of the series which provides 
the dominant behavior. For this, one combines in the corresponding 
alternating series two successive terms, which leads to the factorisation 
of the parameter $m$. Proceeding then as for $F_1$, one finds, as
expected,
\be \lb{3e18b}
F_0(p^2)_{\stackrel{{\displaystyle =}}{p^2\rightarrow -\infty}}
\ \frac{im}{p^2},\ \ \ \ \ m\neq 0.
\ee
In $x$-space, $F_0(r)$ behaves for spacelike $x$ and small $r$ as 
$(m/(4\pi))\ln(1/(\sigma r^2))$.
\par
When $m=0$, the mass dependent terms in the left-hand side of Eqs.
(\rf{3e8}) and (\rf{3e9}) disappear. Eq. (\rf{3e9}) implies that
$\frac{\partial}{\partial r}(\widetilde F_0(r))$ vanishes at $r=0$,
which leads to a constraint on the masses and residues:
\be \lb{3e18}
\sum_{n=1}^{\infty}\,(-1)^nb_nM_n=0,\ \ \ \ \ \ m=0.
\ee
$F_0(0)$ is then a constant. Analyzing the behavior of the 
right hand-side at small $r$, one finds for $F_0(r)$ the following 
behavior: 
$F_0(r)=F_0(0)\Big[1-(\sigma r^2/(4\pi))\ln(1/(\sigma r^2))\Big]+O(r^2)$.
By Fourier transformation to momentum space, the first term 
produces a delta-function that does not contribute to the 
asymptotic behavior; it is the logarithmic term that provides the
dominant term in the asymptotic region \cite{gs} ($r^2=-x^2$, Eq.
(\rf{3e5})):
\be \lb{3e18d}
F_0(p^2)_{\stackrel{{\displaystyle =}}{p^2\rightarrow -\infty}}
-\frac{4i\sigma F_0(x=0)}{(p^2)^2}=
\frac{2i\sigma}{N_c}\frac{\langle\overline\psi\psi\rangle}
{(p^2)^2},\ \ \ \ \ \ m=0,
\ee
where in the second equation we have introduced the one-flavor quark 
condensate\\ $\langle\overline\psi(0)\psi(0)\rangle=-2N_cF_0(x=0)$.
\par  
We have thus shown that expressions (\rf{3e1})-(\rf{3e3}), or 
equivalently (\rf{3e5})-(\rf{3e7}), are solutions of Eq. (\rf{2e3}),
provided the constraints (\rf{3e10})-(\rf{3e12}) are satisfied.
\par
On practical grounds, there remains to solve the latter equations
for the parameters $M_n$ and $\overline m_n$ ($n=1,2,\ldots$).
The properties of those equations are studied in the appendix,
where also details about their numerical resolution are presented. 
We have calculated, for several values of the quark mass,
$m=0.,\ 0.1,\ 1.,\ 5.,\ 20.$, in mass unit of 
$\sqrt{\sigma/\pi}\simeq 240$ MeV, the first 900 values of the 
parameters $M_n$ and $\overline m_n$, with better
accuracy than $10^{-4}$ for the first 400 values. For massless or 
light quarks, the asymptotic values (\rf{3e4}) are reached within a 
few per cent accuracy after a few dozens of $n$s. With increasing quark 
masses, the asymptotic limit is reached more slowly. Approximate formulas 
for the next-to-leading terms in the asymptotic behavior of $M_n$ are 
presented in the appendix. We display in Tables \rf{3t1} and \rf{3t2} a 
few typical values of $M$ and $\overline m$.
\par
\begin{table}[ht]
\bc
\begin{tabular}{|c|c|c|c|c|c|c|c|c|}
\hline
$n$ & 1 & 2 & 3 & 4 & 5 & 6 & 225 & 226 \\
\hline
$m=0$ & 1.161 & 3.043 & 4.301 & 5.287 & 6.127 & 6.868 & 46.96 & 47.06\\
$m=0.1$ & 1.248 & 3.103 & 4.344 & 5.323 & 6.158 & 6.897 & 46.96 & 47.06\\
$m=1$ & 2.053 & 3.706 & 4.823 & 5.745 & 6.536 & 7.246 & 47.03 & 47.14\\
$m=5$ & 5.833 & 7.059 & 7.898 & 8.642 & 9.287 & 9.887 & 47.92 & 48.02\\
$m=20$ & 20.59 & 21.41 & 21.97 & 22.48 & 22.93 & 23.36 & 55.61 & 55.70\\
\hline
\end{tabular}
\ec
\caption{A few values of $M$, in mass unit of $\sqrt{\sigma/\pi}$,
for several values of the quark mass $m$.}
\lb{3t1}
\end{table}
\par 
\begin{table}[ht]
\bc
\begin{tabular}{|c|c|c|c|c|c|c|c|c|}
\hline
$n$ & 1 & 2 & 3 & 4 & 5 & 6 & 225 & 226 \\
\hline
$m=0$ & 1.052 & 0.217 & 0.282 & 0.147 & 0.191 & 0.120 & 
0.022 & 0.021 \\
$m=0.1$ & 0.969 & 0.207 & 0.282 & 0.143 & 0.192 & 0.117 & 
0.022 & 0.021 \\
$m=1$ & 0.553 & 0.145 & 0.257 & 0.110 & 0.189 & 0.094 &
0.023 & 0.019 \\
$m=5$ & 0.179 & 0.056 & 0.143 & 0.051 & 0.125 & 0.048 &
0.025 & 0.016 \\
$m=20$ & 0.049 & 0.016 & 0.047 & 0.016 & 0.046 & 0.016 & 
0.022 & 0.010 \\
\hline
\end{tabular}
\ec
\caption{A few values of $\overline m$, in mass unit of 
$\sqrt{\sigma/\pi}$, for several values of the quark mass $m$.}
\lb{3t2}
\end{table}
\par
We present in Fig. \rf{3f2} the function $iF_0$ for spacelike $p$ and 
in Fig. \rf{3f3} its real part for timelike $p$, and similarly for
the function $iF_1$ in Figs. \rf{3f4} and \rf{3f5}, for $m=0$. Their
imaginary parts (the spectral functions multiplied with $\pi$) have
behaviors that resemble those of the real parts in the timelike region,
except that they are null below $M_1$ and that the infinite singularities
begin just from the right of the $M$s. A relationship between $iF_1(0)$
and $iF_0(0)$ exists; it is presented in the appendix.
\par
\bfg
\vspace*{0.5 cm}
\bc
\input{3f2.tex}
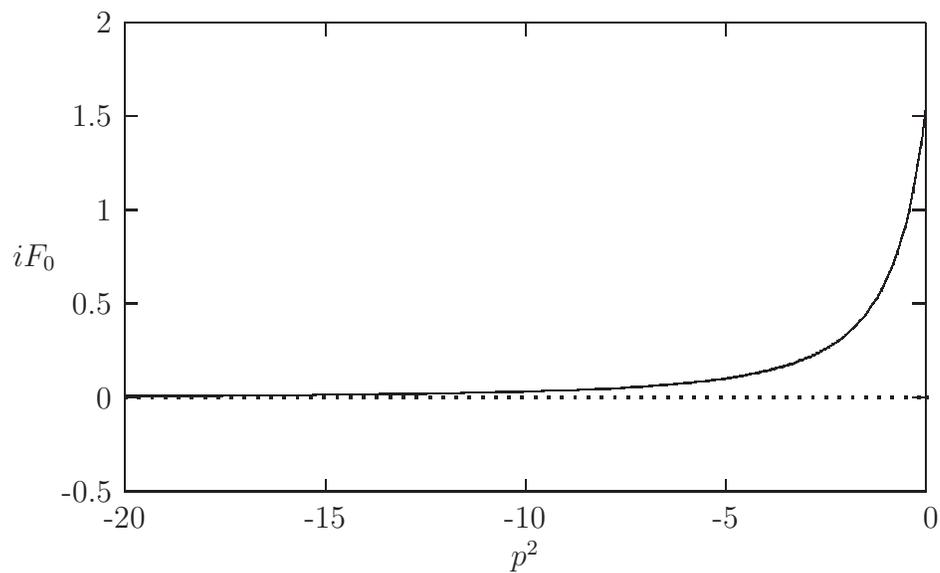
\caption{The function $iF_0$ for spacelike $p$, in mass unit of 
$\sqrt{\sigma/\pi}$, for $m=0$.} 
\lb{3f2}
\ec
\efg
\par
\bfg
\vspace*{0.5 cm}
\bc
\input{3f3.tex}
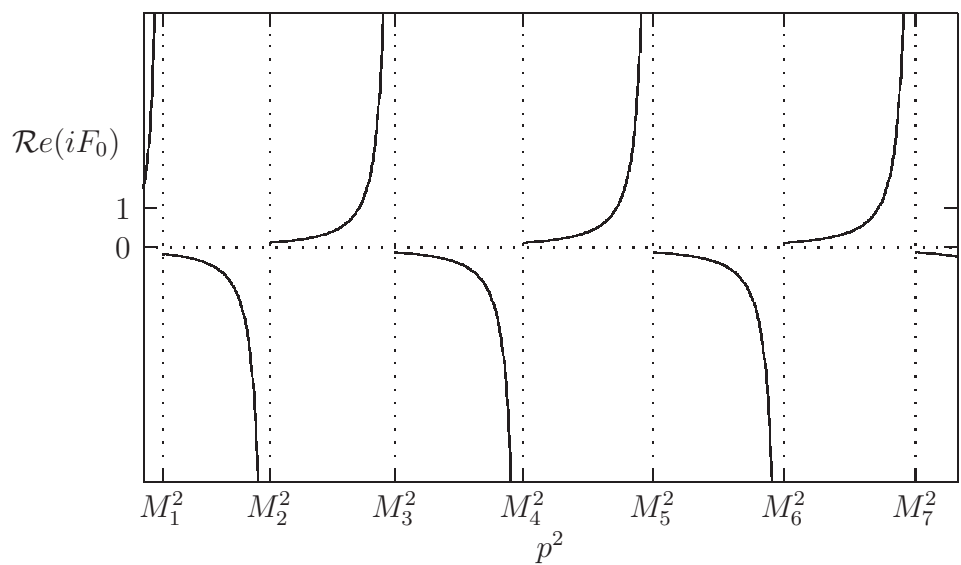
\caption{The real part of the function $iF_0$ for timelike $p$, in mass 
unit of $\sqrt{\sigma/\pi}$, for $m=0$.} 
\lb{3f3}
\ec
\efg
\par
\bfg
\vspace*{0.5 cm}
\bc
\input{3f4.tex}
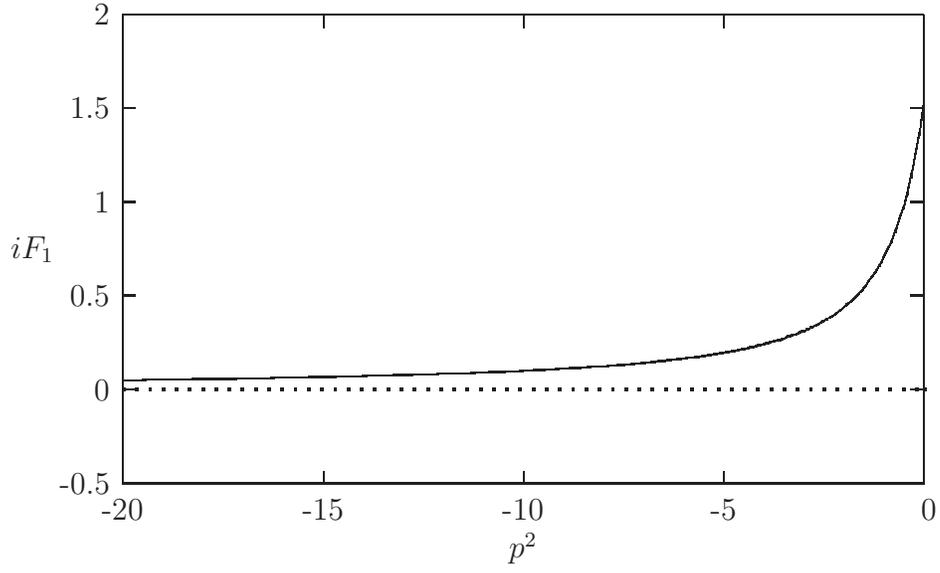
\caption{The function $iF_1$ for spacelike $p$, in mass unit of 
$\sqrt{\sigma/\pi}$, for $m=0$.} 
\lb{3f4}
\ec
\efg
\par
\bfg
\vspace*{0.5 cm}
\bc
\input{3f5.tex}
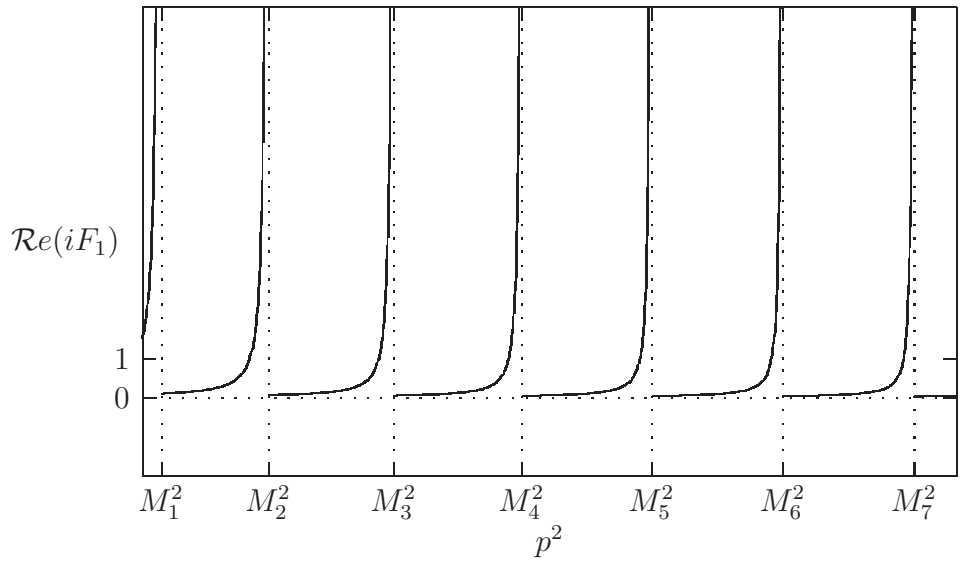
\caption{The real part of the function $iF_1$ for timelike $p$, in mass 
unit of $\sqrt{\sigma/\pi}$, for $m=0$.} 
\lb{3f5}
\ec
\efg
The $x$-space functions $F_0(r)$ and $F_1(r)$ [Eqs. (\rf{3e6})-(\rf{3e7})]
are positive and monotone decreasing functions of $r$ for spacelike $x$. 
For timelike $x$, the two functions are exponentially oscillating.   
\par

\section{Momentum space analysis: the instantaneous limit} \lb{s4}
\setcounter{equation}{0}

A direct resolution of Eq. (\rf{2e3}) in momentum space is complicated
because of the presence of the additional integrations upon the
variables $\lambda$ and $\xi$. Nevertheless, a two-step resolution,
using the analyticity properties resulting from the presence of spectral 
functions, is possible. This section and the following one describe the 
corresponding procedure.
\par
The starting point is based on the observation that Eq. (\rf{2e3})
is quasi-local in $x$, in the sense that in the right-hand side of
the equation $x$ undergoes only dilatation operations. Therefore,
projections of the Green's function on particular subspaces still
give rise to the same equation (for $x\neq 0$). Of particular
interest is the instantaneous limit $x^0=0$. In that case, the variable
$r$ of Eq. (\rf{3e5}) simply becomes the modulus of the space component
$x^1$: $r=|x^1|$. The functions $\widetilde F_1(r)$ and $\widetilde F_0(r)$,
introduced in Eqs. (\rf{3e6}), keep the same definitions and Eqs.
(\rf{3e8}) and (\rf{3e9}) remain unchanged in form with the only 
modification that $r$ has become a one-dimensional variable.  
\par
We pass now to momentum space with one-dimensional Fourier transformation.
The momentum variable $k$ will designate the modulus of the space component
$p_1$ of the two-dimensional vector $p$: $k=|p_1|$. The Fourier transform 
of $\widetilde F_0(r)$ will be defined with the cosine function, while that
of $\widetilde F_1(r)$, which is defined as a derivative of $F_1(r)$
with respect to $r$ [Eq. (\rf{3e6})], with the sine function.
We have the definitions
\bea
\lb{4e1}
& &f_0(k)=2\int_0^{\infty}\,dr\,\cos(kr)\,\widetilde F_0(r),\ \ \  
f_1(k)=2\int_0^{\infty}\,dr\,\sin(kr)\,\widetilde F_1(r),\ \ \   
r=|x^1|,\\
\lb{4e2}
& &F_{0I}(k)=\int_{-\infty}^{\infty}\,\frac{dp_0}{2\pi}\,F_0(p)
=\frac{f_0(k)}{2\pi},\ \ \ 
F_{1I}(k)=\int_{-\infty}^{\infty}\,\frac{dp_0}{2\pi}\,F_1(p)
=\frac{f_1(k)}{2\pi k},\ \ \  k=|p_1|.\nonumber \\
& & 
\eea
\par
Replacement of $\widetilde F_1(r)$ and $\widetilde F_0(r)$ in
Eqs. (\rf{3e8}) and (\rf{3e9}) by their one-dimensional Fourier
transforms allows the integration with respect to the variables
$\lambda$ and $\xi$. One ends up with the equivalent equations for 
$f_1(k)$ and $f_0(k)$:
\bea
\lb{4e3}
k\frac{\partial f_1(k)}{\partial k}
+m\frac{\partial f_0(k)}{\partial k}&=&-\frac{\sigma}{2\pi^2}
\int_0^{\infty}dq\,\Big\{\,(f_1^2(q)+f_0^2(q))\,(\frac{1}{(q+k)^3}
-\frac{1}{(q-k)^3})\nonumber \\
& &\ \ \ -(f_1(q)\frac{\partial f_1(k)}{\partial k}
-f_0(q)\frac{\partial f_0(k)}{\partial k})\,\frac{1}{(q+k)^2}
\nonumber\\
& &\ \ \ +(f_1(q)\frac{\partial f_1(k)}{\partial k}
+f_0(q)\frac{\partial f_0(k)}{\partial k})\,\frac{1}{(q-k)^2}\,
\Big\}-\frac{\sigma}{\pi^2}f_1^2(0)\frac{\partial}{\partial k}
\delta(k),\nonumber \\
& &\\
\lb{4e4}
kf_0(k)-mf_1(k)&=&\frac{\sigma}{2\pi^2}\int_0^{\infty}dq\,\Big\{\,
(f_1(q)f_0(k)+f_0(q)f_1(k))\,\frac{1}{(q+k)^2}
\nonumber\\
& &\ \ \ -(f_1(q)f_0(k)-f_0(q)f_1(k))\,\frac{1}{(q-k)^2}\,\Big\}.
\eea
\par
Equations (\rf{4e3}) and (\rf{4e4}) display singular kernels. Their
effects on the solutions can be studied by expanding the variable
$q$ around $k$ in the functions $f_1(q)$ and $f_0(q)$. It turns out
that Eq. (\rf{4e4}) is finite, while Eq. (\rf{4e3}) is in general
divergent. Finite solutions exist, provided that they satisfy the
constraint $f_1^2+f_0^2=$constant. The value of the constant can be
determined from the inhomogeneous part of Eq. (\rf{2e3}) and the 
property that $f_1$ and $f_0$ should behave asymptotically as in a free 
field theory, i.e., $f_1(k)\rightarrow \pi$ and $f_0(k)\rightarrow 0$ for
$k\rightarrow \infty$. This fixes the constant to $\pi^2$:
\be \lb{4e5}
f_1^2(k)+f_0^2(k)=\pi^2.
\ee
\par
Parametrizing the functions $f_1$ and $f_0$ by means of an angle 
$\varphi(k)$ as
\be \lb{4e6}
f_1(k)=\pi\cos\varphi(k),\ \ \ \ \ f_0(k)=\pi\sin\varphi(k),
\ee
Eq. (\rf{4e3}) reproduces Eq. (\rf{4e4}) in addition to its last term
(the derivative of the delta-function). The compatibility of the
two equations then requires the vanishing of the coefficient of the
latter term, i.e., of $f_1(0)$:
\be \lb{4e7}
f_1(0)=0,\ \ \ \ \mathrm{or},\ \ \ \ \varphi(0)=\frac{\pi}{2}.
\ee
The final equation is then:
\bea \lb{4e8}
k\sin\varphi(k)-m\cos\varphi(k)
&=&\frac{\sigma}{2\pi}\Big\{\,\cos\varphi(k)
\int_0^{\infty}dq\,\sin\varphi(q)\,(\frac{1}{(q+k)^2}+\frac{1}{(q-k)^2})
\nonumber \\
& &\ \ \ \ +\sin\varphi(k)\int_0^{\infty}dq\,
\cos\varphi(q)\,(\frac{1}{(q+k)^2}-\frac{1}{(q-k)^2})\,\Big\}.\nonumber\\
& &
\eea
Let us emphasize at this point that the boundary condition (\rf{4e7}),
which is a necessary condition for the existence of solutions to Eqs.
(\rf{4e3})-(\rf{4e4}), eliminates the free case $\varphi=0$ when $m=0$
from the set of possibilities, which otherwise would fit Eq. (\rf{4e8}). 
This conclusion can also be reached directly in $x$-space, by inspecting 
Eq. (\rf{2e3}). 
\par
Equation (\rf{4e8}) is identical to the one satisfied by the ordinary
quark propagator in its instantaneous limit in the axial gauge
\cite{bg,lwb,knv}. Therefore, the 2PGIQGF and the ordinary quark 
propagator in the axial gauge coincide in their instantaneous limits. 
\par
More generally, Eq. (\rf{4e8}) is analogous to the self-energy 
equation obtained in the instantaneous limit in the four-dimensional 
theory, using Coulomb gluons for the description of the confinement 
mechanism \cite{fgmw,alyopro,ad,aa,lg}. Here, however, as well as in 
the axial gauge of the two-dimensional theory, the ordinary quark 
propagator is a divergent quantity due to the infra-red singularity of 
the gluon-propagator, and is made finite with the aid of a regularization
procedure. It turns out that its instantaneous limit is free of 
divergences and satisfies a finite equation \cite{ad}. This is the reason 
why the interface between the 2PGIQGF, which in general is expected to be 
an infra-red finite quantity, and the ordinary quark propagator in a 
non-covariant gauge, exists only in the instantaneous limit.    
\par
From Eq. (\rf{4e8}), one deduces that asymptotically, as 
$k\rightarrow\infty$, $\sin\varphi$ behaves as $m/k$ when $m\neq 0$
and as $\frac{2\sigma}{k^3}\int_0^{\infty}\frac{dq}{2\pi}\sin\varphi(q)$ 
when $m=0$, in agreement with the results obtained in Eqs. (\rf{3e18b}) 
and (\rf{3e18d}).
\par 
Equation (\rf{4e8}) can be solved with the same methods as its
four-dimensional analogue \cite{ad,ap}. The solution for
$\sin\varphi$, for massless quarks, is presented in Fig. \rf{4f1}. 
\par
\bfg
\vspace*{0.5 cm}
\bc
\input{4f1.tex}
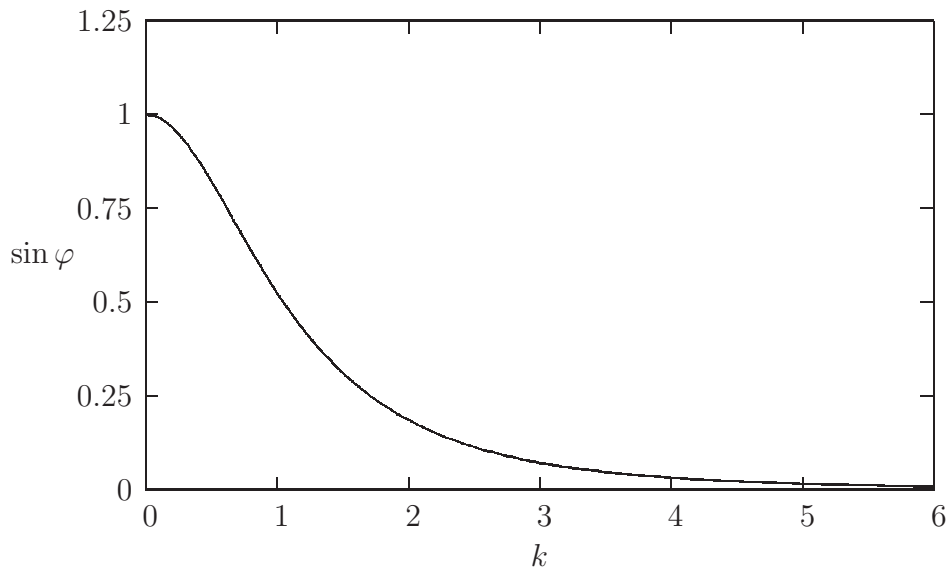
\caption{The function $\sin\varphi$, in mass unit of 
$\sqrt{\sigma/\pi}$, for $m=0$.} 
\lb{4f1}
\ec
\efg
\par

\section{Analyticity properties in the instantaneous limit} \lb{s5}
\setcounter{equation}{0}

The numerical knowledge of the instantaneous limit of the functions 
$F_1(p)$ and $F_0(p)$ does not allow an immediate reconstruction of
the whole covariant functions and in particular of their spectral
functions. Nevertheless, the existence of underlying analyticity 
properties allows us to reach that aim by means of an analytic
continuation. This section is devoted to that procedure.
\par
To simplify the presentation, we shall assume that the series of
spectral functions (\rf{1e2}) sum up, by means of integrations
by parts, into the lowest-order denominator. If the spectral
functions possess strong singularities that forbid the above 
operations, one can bring outside the integrals the appropriate
number of derivation operators. The spectral representation of 
the functions $F_1$ and $F_0$ then becomes:
\be \lb{5e1}
F_1(p^2)=i\int_0^{\infty}ds\,\frac{\rho_1(s)}{(p^2-s+i\varepsilon)},
\ \ \ \ \ F_0(p^2)=i\int_0^{\infty}ds\,\frac{\rho_0(s)}
{(p^2-s+i\varepsilon)}.
\ee
\par
Taking the instantaneous limit of the latter functions [Eqs. (\rf{4e2}) 
and (\rf{4e6})], one finds
\be \lb{5e2}
\cos\varphi(k)=k\int_0^{\infty}ds\,\frac{\rho_1(s)}{\sqrt{k^2+s}},
\ \ \ \ \ \sin\varphi(k)=\int_0^{\infty}ds\,\frac{\rho_0(s)}
{\sqrt{k^2+s}}.
\ee
\par
In principle, the above integral equations could be inverted with the
successive uses of the Hankel transform of order zero and the inverse
Laplace transform \cite{e}. However, the latter transform necessitates 
knowledge of analytic expressions. Also, the direct calculation of the
product of the two transforms gives a singular expression, since the
Bessel function of order zero is an analytic function. These difficulties
can, however, be circumvented with the use of an analytic continuation 
method.
\par
The key observation is that because of the properties of the spectral
functions, assumed to be concentrated on the positive real axis, the 
instantaneous limit functions $\cos\varphi$ and $\sin\varphi$ are 
themselves analytic functions of $k$ in its whole complex plane with 
possible singularities located on the imaginary axis. We are then 
entitled to continue those functions to complex values of $k$ and in 
particular to the imaginary axis.
\par
Let us consider for definiteness the function $\sin\varphi$. One can
express it as the Laplace transform of some other function, which
will be denoted $h_0$:
\be \lb{5e3}
\sin\varphi(k)=\int_0^{\infty}dy\,h_0(y)e^{{\displaystyle -yk}}.
\ee
The exponential function $e^{-yk}$ can be expressed as the Hankel 
transform of order zero of the function $1/\sqrt{k^2+s}$:
\be \lb{5e4}
e^{{\displaystyle -yk}}=\frac{y}{2}\int_0^{\infty}ds\,\frac{1}
{\sqrt{k^2+s}}J_0(y\sqrt{s}),
\ee
where $J_0$ is the Bessel function of order zero.
Replacing in Eq. (\rf{5e3}) the exponential function by its
expression (\rf{5e4}), one arrives at the identification
\be \lb{5e5}
\rho_0(s)=\frac{1}{2}\int_0^{\infty}dy\,h_0(y)yJ_0(y\sqrt{s}).
\ee 
Finally, using for the Bessel function the representation
$J_0(z)=(1/\pi)\int_{-1}^{+1}dte^{izt}/\sqrt{1-t^2}$ and continuing
Eq. (\rf{5e3}) to complex values of $k$, one obtains the formulas
\bea \lb{5e6}
\rho_0(s)&=&\frac{1}{2i\pi}\int_0^{\sqrt{s}}du\,\frac{1}{\sqrt{s-u^2}}
\frac{\partial}{\partial u}\Big(\sin\varphi(-iu)-\sin\varphi(iu)\Big)
\nonumber \\
&=&\frac{2}{2i\pi}\frac{\partial}{\partial s}
\int_0^{\sqrt{s}}du\,\sqrt{s-u^2}
\frac{\partial}{\partial u}\Big(\sin\varphi(-iu)-\sin\varphi(iu)\Big).
\eea
\par
Assuming that the real part of $\sin\varphi$ is an even function on
the imaginary axis and the imaginary part an odd function, Eq. 
(\rf{5e6}) means that the spectral function $\rho_0$ is determined from 
the imaginary part of the function $\sin\varphi$ on the imaginary axis.
\par
A similar calculation as above can be repeated for the function
$\cos\varphi$. One obtains
\bea \lb{5e7}
\rho_1(s)&=&\frac{1}{2\pi}\int_0^{\sqrt{s}}du\,\frac{1}{\sqrt{s-u^2}}
\frac{\partial}{\partial u}\Big[\frac{1}{u}
(\cos\varphi(-iu)+\cos\varphi(iu))\Big]\nonumber \\
&=&\frac{2}{2\pi}\frac{\partial}{\partial s}
\int_0^{\sqrt{s}}du\,\sqrt{s-u^2}
\frac{\partial}{\partial u}\Big[\frac{1}{u}
(\cos\varphi(-iu)+\cos\varphi(iu))\Big].
\eea
The spectral function $\rho_1$ is then determined from the real part
of $\cos\varphi$ on the imaginary axis.
\par
To have an elementary check of the consistency of Eqs. (\rf{5e6}) and 
(\rf{5e7}), let us assume that the spectral functions are null for values 
of $s$ below a positive mass-squared $M^2$. Then, for values of $k$ 
smaller than $M$, one has from Eqs. (\rf{5e2}) the expansions 
$\sin\varphi(k)=1+ak^2+O(k^4)$ and $\cos\varphi(k)=bk+O(k^3)$. The 
analytic continution to the imaginary axis gives: 
$\sin\varphi(\pm ik)=1-ak^2+O(k^4)$, $\cos\varphi(\pm ik)=\pm ibk+O(k^3)$, 
from which one verifies, through Eqs. (\rf{5e6}) and (\rf{5e7}), that 
$\rho_0$ and $\rho_1$, as expected, are null in the considered domain.
\par  
The evaluation of expressions (\rf{5e6}) and (\rf{5e7}) necessitates 
the analytic continuation of the functions $\sin\varphi$ and 
$\cos\varphi$ to the imaginary axis of the complex variable $k$. This 
can be achieved by using the integral equation (\rf{4e8}), in which the 
integrands $\sin\varphi(q)$ and $\cos\varphi(q)$, which lie on the 
positive real axis, will now be assumed numerically known from the 
resolution of the equation for real positive values of $k$. The validity 
of that continuation is justified by the fact that Eq. (\rf{4e8}) is the 
instantaneous limit of covariant scalar equations relative to the 
functions $F_1$ and $F_0$. The singularities of the integrals arise from 
the terms $1/(q-k)^2$, which actually contribute, by combining the various 
terms, as a principal value integrand. (The corresponding definition 
results from the Fourier sine or cosine transforms of the terms of the 
one-dimensional limit of Eq. (\rf{2e3}).) The latter has a well-defined 
meaning as a generalized function \cite{gs}. One has:
\be \lb{5e8}
P\frac{1}{(q-k)}=\frac{1}{q-(k\mp i\varepsilon)}\pm i\pi\delta(q-k).
\ee
[$P$ designates the principal value.] The first term in the right-hand
side of Eq. (\rf{5e8}) is the boundary value of an analytic function 
and can readily be continued to the complex plane. According to the
sign in front of $i\varepsilon$, the continuation will be done to the
lower half-plane (minus sign) or to the upper half-plane (plus sign).
The delta-function can be integrated with respect to $q$ (actually,
preceded by an integration by parts) and yields a term proportional
to $\frac{\partial}{\partial k}\varphi(k)$, which is an analytic 
function of $k$ and can be continued to the complex plane.
\par
We shall mainly be interested by the limit of $k$ to the imaginary axis 
(with a small positive real part) and therefore shall directly consider
this case:
\be \lb{5e9}
k\rightarrow \eta\pm ik_i,\ \ \ \ \eta>0,\ \eta\simeq 0,\ \ \ \ k_i>0.
\ee
[For simplicity, $\eta$ will henceforth be omitted from the formulas.]
With the above definitions, the continuation of Eq. (\rf{4e8}) to the
imaginary axis takes the form
\bea \lb{5e10}
& &\pm ik_i\sin\varphi(\pm ik_i)-m\cos\varphi(\pm ik_i)
+\frac{\sigma}{2}\frac{\partial}{\partial k_i}\varphi(\pm ik_i)
\mp i\frac{\sigma}{2}\varphi'(0)\sin\varphi(\pm ik_i)
\nonumber \\
& &\ \ \ \ \ \ =\cos\varphi(\pm ik_i)\,g_1(k_i)\mp i\sin\varphi(\pm ik_i)\,
g_2(k_i),
\eea
where $\varphi'(0)=\frac{\partial\varphi(k)}{\partial k}|_{k=0}$,
and $g_1$ and $g_2$ are the following integrals:
\bea
\lb{5e11}
g_1(k_i)&=&\frac{\sigma}{\pi}\int_0^{\infty}dq\,
\frac{(q^2-k_i^2)}{(q^2+k_i^2)^2}\,\sin\varphi(q),\\
\lb{5e12}
g_2(k_i)&=&\frac{\sigma}{\pi}\int_0^{\infty}dq\,
\arctan(\frac{k_i}{q})\,\frac{\partial^2}
{\partial q^2}\cos\varphi(q)\nonumber \\
&=&\frac{\sigma}{2}\varphi'(0)+\frac{\sigma}{\pi}k_i\int_0^{\infty}dq\,
\frac{1}{(q^2+k_i^2)}\,\frac{\partial}{\partial q}\cos\varphi(q).
\eea
The first expression of $g_2$ is useful for small values of $k_i$, 
while the second one for large values of $k_i$; one has
$g_2(\infty)=\frac{\sigma}{2}\varphi'(0)$. $g_1$ and $g_2$ are in 
general negative. In magnitude, $|g_1|$ is a monotone decreasing 
function from a finite value at $k_i=0$ down to zero at infinity; 
$|g_2|$ is a monotone increasing function from zero at $k_i=0$ 
up to $\sigma|\varphi'(0)|/2$ at infinity ($\varphi'(0)$ is negative).
\par
The functions $g_1(k_i)$ and $g_2(k_i)$ being known from the resolution
of Eq. (\rf{4e8}), we find that the continuation of the latter equation
to the imaginary axis has yielded a first-order differential equation in
$\varphi$, which can be analyzed in turn.
\par
Decomposing $\varphi$ into a real and an imaginary part, 
$\varphi(k)=\varphi_r(k)+i\varphi_i(k)$, Eqs. (\rf{5e10}) decompose into
two real equations:
\bea
\lb{5e13}
& &\pm k_i\sin\varphi_r(\pm ik_i)\cosh\varphi_i(\pm ik_i)
+m\sin\varphi_r(\pm ik_i)\sinh\varphi_i(\pm ik_i)\nonumber \\
& &\ \ \ \ +\frac{\sigma}{2}\frac{\partial}{\partial k_i}\varphi_i(\pm ik_i)
\mp \frac{\sigma}{2}\varphi'(0)\sin\varphi_r(\pm ik_i)
\cosh\varphi_i(\pm ik_i)
\nonumber \\
& &\ \ \ =-\sin\varphi_r(\pm ik_i)\sinh\varphi_i(\pm ik_i)\,g_1(k_i)
\mp\sin\varphi_r(\pm ik_i)\cosh\varphi_i(\pm ik_i)\,g_2(k_i),\nonumber \\
& &\\
\lb{5e14}
& &\mp k_i\cos\varphi_r(\pm ik_i)\sinh\varphi_i(\pm ik_i)
-m\cos\varphi_r(\pm ik_i)\cosh\varphi_i(\pm ik_i)\nonumber \\
& &\ \ \ \ +\frac{\sigma}{2}\frac{\partial}{\partial k_i}\varphi_r(\pm ik_i)
\pm\frac{\sigma}{2}\varphi'(0)\cos\varphi_r(\pm ik_i)
\sinh\varphi_i(\pm ik_i)
\nonumber \\
& &\ \ \ =\cos\varphi_r(\pm ik_i)\cosh\varphi_i(\pm ik_i)\,g_1(k_i)
\pm\cos\varphi_r(\pm ik_i)\sinh\varphi_i(\pm ik_i)\,g_2(k_i).\nonumber \\
& &
\eea
The boundary values to be imposed are $\varphi_r(0)=\pi/2$ and 
$\varphi_i(0)=0$, i.e., $\varphi(0)=\pi/2$ [Eq. (\rf{4e7})].
\par
The above equations can be solved starting from a domain containing
the origin, to implement the boundary conditions. It is found that 
the singularities of the functions $\sin\varphi$ and $\cos\varphi$ on 
the imaginary axis are represented by an infinite set of simple poles 
with locations at $k=\pm iM_n$, $n=1,2,\ldots\,$. Considering for instance 
the positive imaginary axis, one can divide it into domains, each one
lying between two successive poles, the first domain lying between the
origin and the first pole. In the first domain, one has $\varphi_r=\pi/2$;
in the second one, $\varphi_r=-\pi/2$; in the third one, 
$\varphi_r=\pi/2$ (modulo $2\pi$); $\varphi_r$ thus alternates between 
the constant values $\pm\pi/2$ when passing from one domain to the other.
\par
In each domain, $\varphi_i$ is a monotone increasing or decreasing
function of $k_i$ (taking into account the signs and values of the 
functions $g_1$ and $g_2$). In the first domain, the solution for 
$\varphi_i$ is given by the equation
\bea \lb{5e15}
\sigma\arctan(\tanh(\frac{\varphi_i(k_i)}{2}))&=&\int_0^{k_i}dk_i'
\Big(-k_i'+\frac{\sigma}{2}\varphi'(0)-g_2(k_i')
-(m+g_1(k_i'))\tanh\varphi_i(k_i')\Big),\nonumber \\
& &
\eea
which can be solved by iteration. [We henceforth omit in the argument of
$\varphi_i$ the factor $i$.] The position of the singularity $M_1$
corresponds to that value of $k_i$ for which the right-hand side reaches
the value $\sigma\pi/4$; at that position, $\varphi_i$ becomes
logarithmically infinite. One has then the equation
\be \lb{5e16}
\int_0^{M_1}dk_i'\Big(k_i'-\frac{\sigma}{2}\varphi'(0)+g_2(k_i')
+(m+g_1(k_i'))\,\tanh\varphi_i(k_i')\Big)=\sigma\frac{\pi}{4}.
\ee
\par
In the $(n+1)$th domain ($n\geq 1$), $M_n<k_i<M_{n+1}$, the corresponding 
equations are:
\bea \lb{5e17}
& &\sigma\arctan(\tanh(\frac{\varphi_i(k_i)}{2}))=
(-1)^n\Big\{\sigma\frac{\pi}{4}\nonumber \\
& &\ \ \ \ \ +\int_{M_n}^{k_i}dk_i'
\Big(-k_i'+\frac{\sigma}{2}\varphi'(0)-g_2(k_i')
-(m+g_1(k_i'))\,\tanh\varphi_i(k_i')\Big)\Big\},\nonumber \\
& &\\
\lb{5e18}
& &\int_{M_n}^{M_{n+1}}dk_i'\Big(k_i'-\frac{\sigma}{2}\varphi'(0)+g_2(k_i')
+(m+g_1(k_i'))\,\tanh\varphi_i(k_i')\Big)=\sigma\frac{\pi}{2}.
\eea
\par
The residues of $\sin\varphi$ and $\cos\varphi$ at the position of their
poles are found by expanding in the various expressions $k_i$ around $M_n$.
One finds:
\bea 
\lb{5e19}
& &\sin\varphi(\pm ik_i)_{\stackrel{{\displaystyle \simeq}}{k_i\simeq M_n}}
(-1)^{(n+1)}\frac{\sigma}{2}\frac{1}{(M_n+\overline m_n+(-1)^nm)}
\frac{1}{(M_n-k_i\pm i\varepsilon)},\\
\lb{5e20}
& &\cos\varphi(\pm ik_i)_{\stackrel{{\displaystyle \simeq}}{k_i\simeq M_n}}
\pm i\frac{\sigma}{2}\frac{1}{(M_n+\overline m_n+(-1)^nm)}
\frac{1}{(M_n-k_i\pm i\varepsilon)},\\
\lb{5e21}
& &\overline m_n=-\frac{\sigma}{2}\varphi'(0)+g_2(M_n)+(-1)^ng_1(M_n).
\eea
The real parts of $\sin\varphi$ and $i\cos\varphi$ for $k$ on the
positive imaginary axis are shown in Figs. \rf{5f1} and \rf{5f2}. They
are even and odd functions of $k_i$, respectively.
\par
\bfg
\vspace*{0.5 cm}
\bc
\input{5f1.tex}
\caption{The real part of the function $\sin\varphi$ on the positive
imaginary axis.} 
\lb{5f1}
\ec
\efg
\par
\bfg
\vspace*{0.5 cm}
\bc
\input{5f2.tex}
\caption{The real part of the function $i\cos\varphi$ on the positive
imaginary axis.} 
\lb{5f2}
\ec
\efg
\par
The spectral functions $\rho_0$ and $\rho_1$ are obtained from
Eqs. (\rf{5e6}) and (\rf{5e7}). One finds:
\bea
\lb{5e22}
& &\rho_0(s)=\frac{\sigma}{2}\sum_{n=1}^{\infty}(-1)^n
\frac{M_n}{(M_n+\overline m_n+(-1)^nm)}\frac{\theta(s-M_n^2)}
{(s-M_n^2)^{3/2}},\\
\lb{5e23}
& &\rho_1(s)=-\frac{\sigma}{2}\sum_{n=1}^{\infty}
\frac{1}{(M_n+\overline m_n+(-1)^nm)}\frac{\theta(s-M_n^2)}
{(s-M_n^2)^{3/2}}.
\eea
\par
The functions $\sin\varphi$ and $\cos\varphi$ on the positive real
axis are reconstituted from Eqs. (\rf{5e2}). [Since the $\rho$s have
strong singularities at the various thresholds, one has to use for 
them the first version of equations (\rf{5e6}) and (\rf{5e7}) and make
an interchange in the order of integrations between $u$ and $s$.]
One obtains:
\bea
\lb{5e24}
& &\sin\varphi(k)=-\sigma\sum_{n=1}^{\infty}(-1)^n
\frac{M_n}{(M_n+\overline m_n+(-1)^nm)}\frac{1}{(k^2+M_n^2)},\\
\lb{5e25}
& &\cos\varphi(k)=\sigma k\sum_{n=1}^{\infty}
\frac{1}{(M_n+\overline m_n+(-1)^nm)}\frac{1}{(k^2+M_n^2)}.
\eea
As a check, one verifies that the curve of $\sin\varphi(k)$ obtained
from Eq. (\rf{5e24}) with the numerical values of the mass parameters
$M_n$ and $\overline m_n$ coincides with that obtained from the direct
resolution of Eq. (\rf{4e8}) [Fig. \rf{4f1}].
\par
One can go further, by transforming the informations coming from
the instantaneous limit into an infinite set algebraic equations.
First, going back to the defining equations of the parameters 
$\overline m_n$ [Eqs. (\rf{5e21})] and calculating now the functions
$g_1$ and $g_2$ with the use of the expressions (\rf{5e24}) and (\rf{5e25})
of $\sin\varphi$ and $\cos\varphi$ on the real axis, one finds the set
of equations (\rf{3e17}) allowing the calculation of the $\overline m$s
for known $M$s.
\par
Second, using expressions (\rf{5e24}) and (\rf{5e25}) in Eq. (\rf{4e8}),
one finds, as consistency conditions Eqs. (\rf{3e10}) and (\rf{3e12}).
Equation (\rf{3e11}) is implicit here, since odd-index and even-index 
parameters succeed to each other. [Their definition is not a mere 
artifact, their defining equations being different, cf. Eqs. (\rf{5e17})
and (\rf{5e21}).] The equation $\sin^2\varphi+\cos^2\varphi=1$, which is
a non-trivial constraint for the expressions (\rf{5e24})-(\rf{5e25}), leads
also to the previous consistency relations. One thus finds at the end the 
same algebraic equations as those found in the $x$-space analysis.  
\par
Finally, the expressions (\rf{5e23}) and (\rf{5e22}) of the spectral 
functions allow us to reconstruct the completely covariant functions
$F_1$ and $F_0$, given in Eqs. (\rf{3e2}) and (\rf{3e3}).
\par
This completes the study and determination of the quark Green's 
function in momentum space.       
\par

\section{The quark condensate} \lb{s6}
\setcounter{equation}{0}

A quantity of interest is the quark condensate, which, in the 
one-flavor case and for massless quarks, has the expression 
[Eq. (\rf{3e5})]
\be \lb{6e1}  
\langle\overline \psi(0)\psi(0)\rangle =
-\mathrm{tr}_{c,sp}S(m=0,x)\Big|_{x=0}=-2N_cF_0(m=0,x=0),
\ee
$\mathrm{tr}_{c,sp}$ meaning the trace operation taken on color and
spinor indices. This quantity, being the vacuum average of a local
gauge invariant operator, is also calculable from the ordinary quark
propagator and does not need for its evaluation the recourse to 
gauge invariant Green's functions. Nevertheless, it is interesting to 
compare, as a check, its value obtained in the present approach to 
that obtained in previous approaches.
\par
The quark condensate is calculated from the series expression of the 
scalar function $F_0(x)$ [Eqs. (\rf{3e6}), (\rf{3e7}) and (\rf{3e4a})] 
for $m=0$:
\be \lb{6e2}
\langle\overline \psi(0)\psi(0)\rangle =\frac{2N_c\sigma}{4}
\sum_{n=1}^{\infty}(-1)^n\frac{1}{(M_n+\overline m_n)}\simeq -293\ 
\mathrm{MeV},
\ee
the numerical value being obtained with $N_c=3$ and $\sqrt{\sigma}=425$ 
MeV. [The asymptotic tail of the series is evaluated using the
asymptotic expression of $M_n$ and neglecting the $\overline m$s.]
\par
The above numerical value is in agreement with that obtained in the
axial gauge from the integral of the function $\sin\varphi$ for massless 
quarks \cite{lwb}  ($\langle\overline \psi\psi\rangle \simeq -295$ MeV)
and with that obtained in the light-cone gauge using operator product 
expansion methods \cite{z,blt} 
($\langle\overline \psi\psi\rangle=-N_c\sqrt{\sigma/(6\pi)}\simeq -294$ 
MeV).
\par

\section{Conclusion} \lb{s7}
\setcounter{equation}{0}

The study of the gauge invariant quark Green's function in 
two-dimensional QCD in the large-$N_c$ limit by means of the
integrodifferential equation it satisfies has allowed us 
to determine the properties of its spectral functions. Those
functions directly probe the contributions of quarks and gluons
in a sector where no color-singlet states can contribute. 
The main results can be summarized as follows.
\par
1) The spectral functions are infra-red finite and lie on the positive 
real axis of the variable $p^2$ in Minkowski space ($p$ timelike). No 
singularities in the complex plane or on the negative real axis have been 
found. This feature underlines the fact that quarks and gluons contribute 
to the spectral functions like physical particles with positive energies.
\par
2) The singularities of the spectral functions are represented by an
infinite number of threshold singularities, characterized by positive 
real masses $M_n$ ($n=1,2,\ldots$) extending from a lowest positive
value up to infinity. The corresponding singularities near the thresholds
are of the type $(p^2-M_n^2)^{-3/2}$ and therefore are stronger than 
simple pole singularities, a feature which reflects the difficulty in 
observation of quarks as free particles.
\par
3) The threshold masses $M_n$ remain positive in the limit of massless 
quarks and maintain the scalar part of the Green's function at a non-zero 
value. 
\par
The present method of investigation can also be applied to 
four-dimensional QCD, in which case, however, several difficulties
arise. First, on practical grounds, the integrodifferential equation of 
the gauge invariant Green's function can no longer be solved in an exact
way, because of the presence of an infinite number of kernels and of the
different contributions of short- and large-distance interactions.
Second, the presence of short-distance interactions induces divergences
which necessitate an appropriate treatment through renormalization.  
Third, the Wilson loop averages are no longer known exactly.
\par
Another domain of investigation is the four-point gauge invariant Green's
function, which is naturally related to the bound state equation of mesons. 
\par

\vspace{0.25 cm}
\noindent
\textbf{Acknowledgements}
\par
I thank Nora Brambilla and Antonio Vairo for stimulating discussions 
on the subject and the hospitality at the Technische Universit\"at
M\"unchen. I also thank Antonio Pineda for providing me references in
two-dimensional QCD. This work was supported in part by the EU network 
FLAVIANET, under Contract No. MRTN-CT-2006-035482, and by the European 
Community Research Infrastructure Integrating Activity ``Study of 
Strongly Interacting Matter'' (acronym HadronPhysics2, Grant Agreement 
n. 227431), under the Seventh Framework Programme of EU.
\par

\appendix
\renewcommand{\theequation}{\Alph{section}.\arabic{equation}}

\section{Determination of the spectral function parameters}
\lb{ap1}
\setcounter{equation}{0}

This appendix is devoted to the determination of the threshold 
masses and factors that appear in the spectral functions. In the 
first part, we study the mathematical properties of the corresponding 
equations; in the second part, the asymptotic behaviors of the mass 
parameters $M_n$ and $\overline m_n$ for large $n$ are found and in 
the third part, details of the numerical resolution of the equations 
are given.
\par

\subsection{Mathematical properties} \lb{ap11} 

We notice that Eqs. (\rf{3e10}) and (\rf{3e11}) separate into
two complementary sets according to the parity of the index $n$:
\bea
\lb{3e13a}
& &M_{2n+1}=m+\sum_{q=0}^{\infty}\,\frac{b_{2q+2}}
{(M_{2n+1}-M_{2q+2})^2},\ \ \ \ n=0,1,2,\ldots\,,\\
\lb{3e14a}
& &1=-2\sum_{q=0}^{\infty}\,\frac{b_{2q+2}}
{(M_{2n+1}-M_{2q+2})^3},\ \ \ \ n=0,1,2,\ldots\,,\\
\lb{3e13b}
& &M_{2n+2}=-m+\sum_{q=0}^{\infty}\,\frac{b_{2q+1}}
{(M_{2n+2}-M_{2q+1})^2},\ \ \ \ n=0,1,2,\ldots\,,\\
\lb{3e14b}
& &1=-2\sum_{q=0}^{\infty}\,\frac{b_{2q+1}}
{(M_{2n+2}-M_{2q+1})^3},\ \ \ \ n=0,1,2,\ldots\,.
\eea
The odd-index $M$s are thus determined by equations where the
even-index $M$s play the role of background data. The roles of 
odd-index and even-index $M$s are interchanged in the complementary 
sets of equations and $m$ replaced by $-m$. 
\par
To analyze the constraints imposed by Eqs. (\rf{3e13a})-(\rf{3e14b}),
we may begin by considering Eqs. (\rf{3e13a}) as representing the 
intersection condition of two curves. Let us define the two functions
\be \lb{3e15}
f_{+1}(x)\equiv\sum_{q=0}^{\infty}\frac{b_{2q+2}}
{(x-M_{2q+2})^2},\ \ \ \ \ f_{+2}(x)\equiv x-m,
\ee 
for $x$ real and where the even-index $M$s are assumed to be known.
Then, the solutions of Eqs. (\rf{3e13a}) represent the intersection
points of the straight line $f_{+2}$ with the curve $f_{+1}$. We 
further notice that the function $f_{+1}$ has an infinite number of 
double poles at the even-index masses with positive residues. Any 
solution $M_{2n+1}$ should lie between two successive even-index poles. 
However, in general, the equality $f_{+2}=f_{+1}$ might not lead to such 
a solution. First, the straight line $f_{+2}$ might not intersect at all 
the curve $f_{+1}$, in which case no solution would exist. Or, second, it 
might cut $f_{+1}$ at two different points between two successive 
even-index poles. In that case, we would meet instability, since, 
considering then the complementary sets of equations with the roles of 
odd and even indices interchanged, and with the number of odd-index 
solutions doubled, we would generate new solutions of even-index $M$s 
(doubled in number), and so forth. We therefore guess that the only stable 
and acceptable solutions to Eqs. (\rf{3e13a}) are those corresponding to 
the situation where the straight line $f_{+2}$ is tangent to the curve 
$f_{+1}$ between two successive even-index poles. 
This is precisely the content of Eqs. (\rf{3e14a}). Written as functions 
of $x$, they take the form
\be \lb{3e16}
\frac{df_{+2}(x)}{dx}=\frac{df_{+1}(x)}{dx},
\ee
indicating that the straight line $f_{+2}$ should be tangent to the curve
$f_{+1}$ at the positions of the solutions (cf. Fig. \rf{3f1}). Of course, 
this condition could not be satisfied if the even-index $M$s and 
$\overline m$s were arbitrary. The even-index $M$s should themselves be 
solutions of Eqs. (\rf{3e13b}) and (\rf{3e14b}) ensuring the final 
consistency of all equations.
\par
\bfg
\vspace*{0.5 cm}
\bc
\input{3f1.tex}
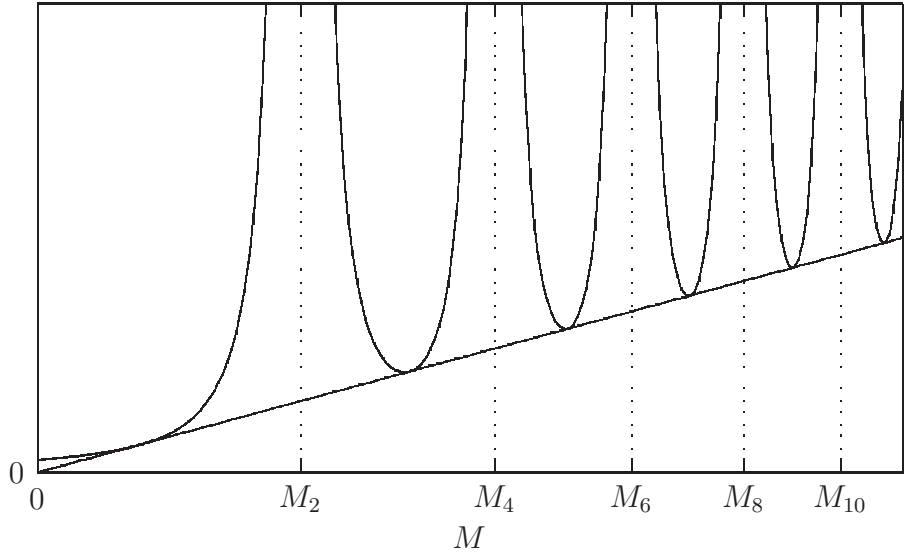
\caption{Graphic representation of Eqs. (\rf{3e13a}) and (\rf{3e14a})
for $m=0$. The solutions, for odd-index parameters, for given even-index 
parameter solutions, correspond to the contact points of the straight line 
with the curves.}
\lb{3f1}
\ec
\efg
While the geometrical interpretation of Eqs. (\rf{3e13a})-(\rf{3e14b})
is very suggestive of the existence of acceptable solutions, one might
still question the mutual compatibility of the latter equations. 
Assuming that the mass parameters $\overline m$, or equivalently the
residues $b$, are mainly determined from Eqs. (\rf{3e12}) or (\rf{3e17}),
one notices that the system of equations (\rf{3e13a})-(\rf{3e14b})
represent four sets of equations for two sets of unknowns, $M_{2n+1}$
and $M_{2n+2}$, $n=0,1,2,\dots\,$. It is then necessary to check the 
compatibility of the above equations.
\par
To that end we define two meromorphic functions $h_+(z)$ and $h_-(z)$:
\bea
\lb{ae13}
h_+(z) &\equiv& z-m-\sum_{q=0}^{\infty}\frac{b_{2q+2}}
{(z-M_{2q+2})^2},\\
\lb{ae14}
h_-(z) &\equiv& z+m-\sum_{q=0}^{\infty}\frac{b_{2q+1}}
{(z-M_{2q+1})^2}.
\eea
The singularities of $h_+$ are double poles located at the even-index
masses, while those of $h_-$ are double poles located at the odd-index 
masses. The residues $b_n$ are defined in Eqs. (\rf{3e4a}) and are 
assumed to be finite quantities. Equations (\rf{3e13a}) take the form 
$h_+(M_{2n+1})=0$, $n=0,1,2,\ldots\,$, which means that $h_+$ has zeros 
at the positions of the poles of $h_-$. Equations (\rf{3e14a}) take the 
form $\frac{dh_+(z)}{dz}\Big|_{z=M_{2n+1}}=0$, which means that the zeros
of $h_+$ are double. Therefore, the product $h_+h_-$ is finite
at the poles of $h_-$. A similar argument, using now Eqs. (\rf{3e13b}) 
and (\rf{3e14b}), also holds at the poles of $h_+$. In conclusion,
the product $h_+(z)h_-(z)$ is an analytic function in the whole complex
$z$-plane. According to Liouville's theorem \cite{ww}, it is a polynomial
of second degree:
\be \lb{ae15}
h_+(z)h_-(z)=z^2-m^2+\frac{2\sigma}{\pi},
\ee
where the right-hand side has been determined from the asymptotic
behaviors of $h_+$ and $h_-$. It is to be emphasized that the formal 
calculation of the product of the series (\rf{ae13}) and (\rf{ae14}) 
by exchanging the orders of summation in the double sum would miss 
the last constant term in the right-hand side of Eq. (\rf{ae15}); this 
is due to the weak convergence of the individual series. 
\par
Equation (\rf{ae15}) indicates that if $h_+$ is known, then $h_-$ is 
completely determined from it. It thus provides the proof of compatibility 
of Eqs. (\rf{3e13a})-(\rf{3e14b}). For suppose that Eqs. (\rf{3e13a}) and
(\rf{3e14a}) are considered on formal grounds as two independent sets 
of equations determining the two sets of unknowns $M_{2n+1}$ and
$M_{2n+2}$. Then the two remaining equations (\rf{3e13b}) and
(\rf{3e14b}) would not bring any new information or constraint 
since their content is already summarized in Eq. (\rf{ae15}).   
On practical grounds, it is still preferable to determine by means
of iterative procedures the odd-index masses from Eqs. (\rf{3e13a})
and (\rf{3e14a}) and the even-index masses from Eqs. (\rf{3e13b}) and
(\rf{3e14b}). 
\par
Relations, involving the residues $b_n$, are obtained by taking in Eq. 
(\rf{ae15}) the limit of $z$ to one of the pole positions of $h_+$ or 
$h_-$. For odd-index poles, for instance, one finds:
\be \lb{ae21}
3b_{2n+1}\sum_{q=0}^{\infty}\frac{b_{2q+2}}{(M_{2n+1}-M_{2q+2})^4}
=M_{2n+1}^2-m^2+\frac{2\sigma}{\pi},
\ee
with a similar relation for the even-index poles. The procedure 
can be continued with higher-order derivatives at the poles, leading 
to new relations for the $b_n$s, but it is not evident whether the
derivation operation inside the series remains still valid at higher 
orders. Equations (\rf{ae15}) and (\rf{ae21}) can be used as 
consistency checks for the solutions that are found.
\par 
We next establish Eq. (\rf{3e17}). The derivation follows similar
lines as above. We define two meromorphic functions $f_+(z)$ and
$f_-(z)$:
\bea 
\lb{ae1}
f_+(z)&\equiv&\sum_{q=0}^{\infty}\,\frac{b_{2q+2}}{(z-M_{2q+2})}
+\sum_{q=0}^{\infty}\,\frac{b_{2q+1}}{(z+M_{2q+1})},\\
\lb{ae2}    
f_-(z)&\equiv&\sum_{q=0}^{\infty}\,\frac{b_{2q+1}}{(z-M_{2q+1})}
+\sum_{q=0}^{\infty}\,\frac{b_{2q+2}}{(z+M_{2q+2})}.
\eea
The function $f_+$ has poles on the positive real axis at even-index
masses and on the negative real axis at minus the odd-index masses.
$f_-$ has analogous properties with the roles of even-index and
odd-index masses interchanged. $f_+$ and $f_-$ satisfy the following 
relation:
\be \lb{ae4}
f_-(z)=-f_+(-z).
\ee
\par
Equations (\rf{3e12}) can be rewritten in the form
\be \lb{ae5}
f_+(M_{2n+1})=0,\ \ \ \ f_-(M_{2n+2})=0,\ \ \ \ \ n=0,1,2,\ldots\,.
\ee
This means that $f_+$ has zeros at the poles of $f_-$ and $f_-$ has
zeros at the poles of $f_+$. Therefore, the product $f_+(z)f_-(z)$
is free of singularities and is analytic in the whole complex $z$-plane.
According to Liouville's theorem \cite{ww}, it is a constant, which
could be determined from the asymptotic behaviors of $f_+$ and $f_-$
at infinity:
\be \lb{ae8}
f_+(z)\,f_-(z)=-\sigma^2.
\ee
[Because of the weak convergence of the series defining the functions 
$f_{\pm}(z)$, it is not possible to obtain the value of the constant by 
formal calculation of the product $f_+(z)f_-(z)$, the latter operation 
being invalid under the exchange of the orders of summation in the double 
series. Asymptotically, the functions $f_{\pm}(z)$ tend to imaginary 
constants.]
\par
It is then possible to calculate the derivatives of $f_+$ and $f_-$
at the position of their zeros. Thus:
\be \lb{ae9}
\frac{df_+(z)}{dz}\Big|_{z=M_{2n+1}}=\sigma^2\lim_{z\rightarrow
M_{2n+1}}\Big(\frac{1}{f_-^2(z)}\frac{df_-(z)}{dz}\Big)=
-\frac{\sigma^2}{b_{2n+1}}.
\ee
Calculating the expression of the left-hand side and then using Eq.
(\rf{3e13a}), we obtain the equation:
\be \lb{ae10}
\frac{\sigma^2}{b_{2n+1}}=M_{2n+1}-m+\sum_{q=0}^{\infty}\,\frac{b_{2q+1}}
{(M_{2n+1}+M_{2q+1})^2},
\ee
which yields
\be \lb{ae11}
b_{2n+1}=\frac{\sigma^2}{M_{2n+1}-m+\sum_{q=0}^{\infty}\,\frac{b_{2q+1}}
{(M_{2n+1}+M_{2q+1})^2}}.
\ee
Comparison with Eq. (\rf{3e4a}) for odd $n$ then provides the expressions
\be \lb{ae12}
\overline m_{2n+1}=\sum_{q=0}^{\infty}\,\frac{b_{2q+1}}
{(M_{2n+1}+M_{2q+1})^2},
\ee
which correspond to the odd-index parts of Eqs. (\rf{3e17}). A similar
derivation can also be done with the even-index parameters.
\par
Actually, the $b_n$s could be defined with a different global 
multiplicative factor than in Eqs. (\rf{3e4a}). The choice done there
is the one which satisfies the correct normalization condition of the
function $F_1$, as emphasized in Sec. \rf{s3}.
\par 
Coming back to the functions $h_{\pm}(z)$ [Eqs. (\rf{ae13}) and 
(\rf{ae14})], it is possible to establish through them a relationship 
between the functions $iF_1(p^2)$ and $iF_0(p^2)$ [Eqs. (\rf{3e1}),
(\rf{3e2}) and (\rf{3e3})] at $p^2=0$. We also include for this analysis  
the series expansions of their instantaneous limits defined in Eqs.
(\rf{4e1}), (\rf{4e2}), (\rf{4e6}), (\rf{4e7}), (\rf{5e24}) and 
(\rf{5e25}), and which involve the function $\varphi(k)$ parametrizing
them ($k$ is the modulus of the space component $p_1$ of the vector $p$). 
Considering the particular values $h_+(0)$ and $h_-(0)$, the derivatives 
$h_+'(0)$ and $h_-'(0)$ and Eq. (\rf{ae15}), one finds the following 
relationships:
\bea
\lb{ae16}
& &\varphi'(0)\Big(-2m+\frac{2\sigma}{\pi}iF_0(0)\Big)=
2-\frac{4\sigma}{\pi}iF_0(0),\\
\lb{ae17}
& &iF_0(0)=\frac{m\pi}{\sigma}+\Big[\frac{\pi^2}{4}\varphi^{\prime 2}(0)+
\frac{\pi}{\sigma}(m^2\frac{\pi}{\sigma}-2)\Big]^{1/2},\\
\lb{ae18}
& &iF_1(0)=\frac{\pi}{2\sigma}-\frac{1}{2}\varphi'(0)
\Big[\frac{\pi^2}{4}\varphi^{\prime 2}(0)+
\frac{\pi}{\sigma}(m^2\frac{\pi}{\sigma}-2)\Big]^{1/2},
\eea
where $\varphi'(0)=\frac{\partial\varphi(k)}{\partial k}\Big|_{k=0}$. 
Elimination of $\varphi'(0)$ yields a relationship between $iF_1(0)$ 
and $iF_0(0)$:
\be \lb{ae19}
iF_1(0)=\frac{\pi}{2\sigma}+\frac{1}{\pi}
\Big(iF_0(0)-\frac{m\pi}{\sigma}\Big)
\Big[(iF_0(0))^2-2iF_0(0)\frac{m\pi}{\sigma}+
\frac{2\pi}{\sigma}\Big]^{1/2}.
\ee
Numerically, one finds, for $m=0$, in mass unit of $\sqrt{\sigma/\pi}$,
from the series expansion
$-\sigma\varphi'(0)=\sum_{n=1}^{\infty}b_n/M_n^2$ [Eq. (\rf{5e25})] and
the values of $M_n$ and $\overline m_n$, 
$\varphi'(0)=-1.325$, which predicts $iF_1(0)=1.512$
and $iF_0(0)=1.528$, in complete agreement with their direct calculation 
from the series (\rf{3e2}) and (\rf{3e3}).
\par
Other relations can also be obtained, like the expressions of the 
functions $\varphi'$ and $\varphi$ in the form of series and 
alternative expressions of the residues $b_n$ in the form of infinite
products of factors. However, these results are not of direct practical
relevance for the present calculations and will not be displayed. 
\par

\subsection{Asymptotic expansions} \lb{ap12}

The asymptotic behaviors of the mass parameters $M_n$ and 
$\overline m_n$ for large $n$ can be studied from Eqs. (\rf{5e18}) and 
(\rf{5e21}). If the quark mass $m$ is itself large, then one has to 
distinguish two different regions: one for which $\sigma\pi n\gg m^2$ 
and one for which $\sigma\pi n\ll m^2$. If $m$ is small or null, then 
the first region may be identified with the region defined by the large 
$n$ values.
\par
We first consider the first region ($k_i^2\sim\sigma\pi n\gg m^2$).
In the integral (\rf{5e18}), $\tanh\varphi_i(k_i)$ varies from
$\pm 1$ to $\mp 1$ when $k_i$ varies between the two mass bounds
$M_n$ and $M_{n+1}$. Therefore, it is the first term of the 
integrand that dominates and yields a recursion relation that can
easily be solved leading to Eq. (\rf{3e4}). [The other contributions
of the integrand vanish at large $k_i$.] The corresponding
behavior of $\overline m_n$ [Eq. (\rf{3e4})] is obtained from the 
contribution of the combination $-\frac{\sigma}{2}\varphi'(0)+g_2(M_n)$
to Eq. (\rf{5e21}), $g_2$ being defined in Eq. (\rf{5e12}).
\par
To obtain an estimate of the next-to-leading term in the large $n$
behavior of $M_n$, it is necessary to specify in more detail the 
structure of the function $\tanh\varphi_i(k_i)$. Using the analytic 
continuation of the functions $\sin\varphi(k)$ [Eq. (\rf{5e24})] and 
$\cos\varphi(k)$ [Eq. (\rf{5e25})] to the imaginary axis, it is easily 
seen that $\tanh\varphi_i(k_i)$ can be expressed in terms of the 
functions $f_+$ and $f_-$ introduced above [Eqs. (\rf{ae1}) and 
(\rf{ae2})]:
\be \lb{ae23}
\tanh\varphi_i(k_i)=\frac{f_+(k_i)+f_-(k_i)}{f_+(k_i)-f_-(k_i)}.
\ee
In order to evaluate the integral of $m\tanh\varphi_i(k_i)$ in Eq.
(\rf{5e18}), one can approximate the functions $f_+$ and $f_-$ by
the contributions of the pole terms in $M_n$ and $M_{n+1}$ and
take into account the contribution of the remainder, whose role
is mainly to ensure the condition that $f_+$ ($f_-$) vanishes at 
the positions of the poles of $f_-$ ($f_+$) [Eqs. (\rf{ae5})], in
the form of effective $k_i$-dependent residues at the above two
poles. Thus, if we are integrating between $M_{2n}$ and $M_{2n+1}$,
we could use the approximations
\be \lb{ae24}
f_+(k_i)\simeq \frac{b_{2n}(k_i-M_{2n+1})}{(M_{2n}-M_{2n+1})
(k_i-M_{2n})},\ \ \ \ 
f_-(k_i)\simeq \frac{b_{2n+1}(k_i-M_{2n})}{(M_{2n+1}-M_{2n})
(k_i-M_{2n+1})},
\ee 
for $M_{2n}\le k_i\le M_{2n+1}$, and similar ones for the interval 
$[M_{2n+1}M_{2n+2}]$. 
\par
With the above approximations, the calculation of the integral
of $\tanh\varphi_i(k_i)$ can be done explicitly. Since we are 
considering the region where $m^2\ll \sigma\pi n$, an expansion in 
$m$, up to linear terms in $m$, can be used. The terms of order zero
in $m$ yield alternating signs when passing from one interval to the 
other and globally do not give significant contributions to the
asymptotic domain. The order one terms in $m$ (multiplied with the
external $m$ factor) can be treated perturbatively with respect to the 
leading $\sigma\pi n$ factors. One finds the following behavior:
\be \lb{ae25}
M_n^2\simeq \sigma\pi (n-n_0)+\Big[(2-\frac{\pi}{2})m^2-
\frac{\sigma}{\pi}\Big]\ln(\frac{n}{n_0})+M_{n_0}^2,\ \ \ \ \ 
n\ge n_0,
\ee
where $n_0$ is the $m$-dependent lower bound of $n$ to be used. We
have also included in the logarithmic term a mass independent contribution 
coming from the combination $-\frac{\sigma}{2}\varphi'(0)+g_2(k_i)$ in
the integrand. The coefficient of the $m^2\ln(n/n_0)$ term is only
approximate.
\par    
We next consider the second region, where $\sigma\pi n\ll m^2$.
This region is of relevance when heavy quarks or nonrelativistic 
expansions are considered. The above approximation for 
$\tanh\varphi_i(k_i)$ can still be used, but now making in the result 
an expansion in $1/m$. The leading term of the expansion cancels the 
contribution of the factor $k_i$ of the integrand and the remaining terms
are suggestive of a separation of the type $M_n=m+m_n$. The 
next-to-leading term contributes with alternating signs and is
globally negligible. Retaining in the remainder the first leading term 
and the next one, the final result is:
\be \lb{ae26}
M_n\simeq m\,\Big[\,1+2\Big(\frac{\sigma n}{3m^2}\Big)^{2/3}
+\frac{16}{9\pi}\Big(\frac{\sigma n}{3m^2}\Big)
\ln\Big(\frac{3m^2}{\sigma n}\Big)\,\Big],\ \ \ \ \ n\le n_0,
\ee
where we have chosen for the upper bound of $n$ the same $n_0$ as
in Eq. (\rf{ae25}), to ensure continuity between the two regions.
The coefficients of the last two terms are approximate.
The expression of the second term in the right-hand side of the above 
equation is typical of a nonrelativistic semiclassical energy of a 
particle of mass $m$ in the presence of a linear confining potential
(with a different coefficient), although we do not have in the present 
problem explicit external forces. The latter result at least 
justifies the validity of nonrelativistic expansions in domains 
where the number $n$ respects the upper bound inequality imposed
by the value of $m$. The mass dependence of the nonrelativistic
term can also be obtained by a direct analysis of Eqs. 
(\rf{3e13a})-(\rf{3e14b}) with the use of the corresponding expressions
of the residues $b_n$.
\par
The domain corresponding to the interval $[0,M_1]$ does not belong
to the asymptotic regions; nevertheless, it is useful to obtain
for $M_1$ an approximate explicit formula. Since that domain is 
concerned only by one mass, the corresponding approximation to be
used for the functions $f_+$ and $f_-$ will be different. Here, one
should keep in $f_-$ its pole term in $M_1$ and in $f_+$ its pole
term in $-M_1$. No further approximation is needed, since in this
case the function $\tanh\varphi_i(k_i)$ correctly vanishes at $k_i=0$.
Explicitly, one has $\tanh\varphi_i(k_i)=-k_i/M_1$ and its integral
can be evaluated. Keeping in Eq. (\rf{5e16}) the factor 
$(k_i+m\tanh\varphi_i(k_i))$ as the integrand, one obtains for $M_1$
the formula
\be \lb{ae27}
M_1^2=\frac{1}{2}\,\Big[\,m^2+\sigma\pi+m\sqrt{m^2+2\sigma\pi}\,\Big],
\ee
which only provides the order of magnitude of $M_1$, but could be 
used in Eq. (\rf{ae25}) when $n_0=1$ (small or null quark mass).
\par
The asymptotic expansions (\rf{ae25}) and (\rf{ae26}) were verified 
in numerical calculations. When $m=20\sqrt{\sigma/\pi}$, the value
$n_0=30$ seems to be optimal. For the other masses that are considered,
$m=0.,\ 0.1,\ 1.,\ 5.$, in unit of $\sqrt{\sigma/\pi}$, $n_0$ can be set
equal to 1 and expansion (\rf{ae25}) can be used with formuula (\rf{ae27}) 
for $M_1$.
\par  

\subsection{Numerical resolution} \lb{ap13}

The reconstitution of the functions $F_1$ and $F_0$ [Eqs. (\rf{3e2}) 
and (\rf{3e3})] and of the functions $\sin\varphi$ and $\cos\varphi$ 
[Eqs. (\rf{5e24}) and (\rf{5e25})] necessitates the knowledge of at 
least a few hundred mass parameters $M_n$. Several difficulties, however, 
are met during the numerical resolution of Eqs. (\rf{3e13a})-(\rf{3e14b}).
\par
Equations (\rf{3e13a})-(\rf{3e14b}) represent two sets of compatible
equations concerning the same unknown quantities (essentially the
$M_n$s). Their compatibility was established when the whole 
contributions of the series were taken into account. Their truncation, 
for computational purposes, introduces a small amount of incompatibility.
Furthermore, each of the series has a different rate of convergence
under iterative procedures. It is therefore not possible to solve
separately and simultaneously these equations without meeting 
instability problems. The best way is to solve one of the compatible
equations, or a combination of the two, and then check the numerical 
validity of the other.
\par
A second difficulty arises from the fact that the distance between 
two successive poles in the series decreases with increasing $q$,
becoming of the order of $\sqrt{\sigma\pi/q}$. This means that the 
corresponding function varies very rapidly between two successive 
poles and small changes in $M_n$ induce big ones in the equations.
\par
To analyze more quantitatively the accuracy of the calculation,
we designate by $e=0$ a generic equation of the type of Eqs.
(\rf{3e13a}) and (\rf{3e13b}), by $e'=0$ its derivative equation,
of the type of Eqs. (\rf{3e14a}) and (\rf{3e14b}), and by $x$ one 
of the masses $M_n$. Using Newton's method of iteration \cite{ap}, 
equation $e'=0$ leads to the following iterative relation: 
$x^{(i+1)}=x^{(i)}-\frac{e^{'(i)}}{e^{''(i)}}$, where $e''$ is the 
second-order derivative of the function $e$ with respect to $x$ ($M_n$)
and $i$ is the order of iteration. $e''$ is negative and for large $n$ 
its modulus reaches values of the order of $10^4-10^6$ (in unit of 
$\sqrt{\sigma/\pi}$). In these domains, the rate of convergence of the 
iteration is very slow and violations of the equation $e'=0$ by factors 
of unity or 10 actually correspond to very small departures of $M_n$ from 
its exact value. The analysis of the equation $e=0$ is more complicated, 
since here the first derivative of $e$ is itself null or small and one 
needs to continue the expansion up to the second-order derivative.
\par
A third difficulty arises from the weak convergence of the series
of Eqs. (\rf{3e13a})-(\rf{3e14b}). It requires a detailed treatment
of the asymptotic tails. 
\par
For the numerical resolution, we have used the combination
$e-e'\sqrt{\sigma/\pi}$ of the equations $e=0$ and $e'=0$ and 
checked after each iteration the separate validity of each of them.
We have considered 900 active points $M_n$ and similarly for the
$\overline m_n$s determined simultaneously through Eqs. 
(\rf{3e17}). As input values, the asymptotic expansions (\rf{ae25})
and (\rf{ae26}) are used. For the asymptotic regions of the various
series, 700 background or passive points, from 900 to 1600, are
considered using the asymptotic formulas (\rf{ae25}) and (\rf{ae26})
with a global scaling factor adjusted after each iteration by 
continuity at the frontier point $n=900$. Beyond that region,
the contribution of the series is approximated by an integral.   
High precision on the results is reached after a number of
iterations that varies between $10^4$ and $5\times 10^4$. The
accuracy of the results concerning the first 400 points is
better than $10^{-4}$ and slightly decreases for the remaining 
points. The uncertainties are mainly due to the contributions
of the asymptotic tails. Finally, various consistency relations 
have been checked: equations (\rf{3e12}), (\rf{ae21}), (\rf{ae15})
(for negative values of $z$), and the constraint 
$\sin^2\varphi(k)+\cos^2\varphi(k)=1$.
\par

\end{document}